\documentclass{article}
\usepackage{cite}
\usepackage{graphicx}
\usepackage{listings}
\usepackage{xcolor}

\title{A simple and fast C++ thread pool implementation capable of running task graphs}
\author{Dmytro Puyda\\dpuyda@gmail.com}
\date{}

\begin{document}
\maketitle
\begin{abstract}
In this paper, the author presents a simple and fast C++ thread pool implementation capable of running task graphs.
The implementation is publicly available on GitHub, see https://github.com/dpuyda/scheduling.
\end{abstract}

\section{Introduction}
Multithreading plays an important role in modern software development. When used wisely, adding more
threads can significantly improve the CPU performance of your application. However, it is well known that
using more threads does not always mean getting better performance. There are at least two common
issues that can happen in multi-threaded software. Firstly, when the number of threads exceeds the
capabilities of your hardware, context switching can have a dramatic impact on the performance of your
application. Secondly, creating and destroying threads frequently can have significant performance
overhead.

To overcome these issues, thread pools are widely used. Instead of creating and destroying threads directly,
developers submit tasks to a thread pool instance. A thread pool typically creates a specified number of
worker threads running in the background. When there are no tasks, the worker threads are idle. When a task
is submitted, one of the available worker threads eventually picks up the task and executes it. If all worker threads
are busy executing other tasks, the new task remains in a task queue until one of the worker threads becomes
available.

As of C++20, there is no thread pool in the ISO standard. However, there are a variety of libraries
providing thread pool implementations that can be used in production. To name just a few: Intel TBB \cite{TBB},
Boost.Asio \cite{BoostAsio}, Taskflow \cite{Taskflow,TaskflowPaper}, CGraph \cite{CGraph}, BS::thread\_pool
\cite{BSThreadPool,BSThreadPoolPaper}, etc. Still, many commercial projects use their own thread pool and task
system implementations.

In this paper, the author suggests a minimalistic and fast work-stealing thread pool implementation capable
of running task graphs. Benchmarks comparing the suggested implementation with Taskflow are provided.
The suggested implementation uses C++20 but can be updated to comply with older versions of the
standard if needed.

The solution suggested in this paper has the following advantages:
\begin{itemize}
\item It is fast and developed with performance in mind. See the benchmark results below for details.
\item It is simple, short, and minimalistic. At the time of writing this paper, the solution consists of less
than one thousand lines of C++ code. New features can be added easily if required.
\item It does not use third-party dependencies and relies only on the C++20 ISO standard.
\end{itemize}
A disadvantage of the suggested solution is that it is not header-only.

\section{Implementation details}
In this section, we describe some implementation details of the suggested solution.

\subsection{Chase-Lev deque}
The idea of work-stealing queues is to provide each worker thread with its own task queue to reduce thread
contention. When a task is submitted, it is pushed to one of the queues. The thread owning the queue can
eventually pick up the task and execute it. If there are no tasks in the queue owned by a worker thread, the
thread attempts to steal a task from another queue.

Work-stealing queues are typically implemented as lock-free deques. The owning thread pops elements at one
end of the deque, while other threads steal elements at the other end.

Implementing a work-stealing deque is not an easy task. The Chase-Lev deque \cite{Chase,Minh} is one of the
most commonly used implementations of such a deque. The reference C11 and ARMv7 implementations, as well as the proof
of correctness of the ARMv7 code, are given in \cite{Minh}. However, \cite{Minh} does not include the proof
of correctness of its C11 implementation. Corrections of the C11 implementation provided in \cite{Minh} were suggested
\cite{Norris,Ou}. Later, a few proofs of correctness were given (see, e.g., \cite{Mutluergil,Choi}). Frameworks
for automatic inference and validation of memory fences have also been used to validate Chase-Lev deque
implementations (e.g., \cite{Ou,Kuperstein}).

The original C11 implementation of the Chase-Lev deque \cite{Minh}, as well as the updated implementation
\cite{Ou}, uses atomic thread fences without associated atomic operations. When compiling with a thread
sanitizer, GCC 13 issues a warning saying that ‘atomic\_thread\_fence’ is not supported with ‘-fsanitize=thread’.
Thread sanitizers may produce false positives when atomic thread fences are used. For example, when using the
Taskflow implementation of the work-stealing deque, the thread sanitizer detects data races in the solution
suggested in this paper. The Taskflow implementation of the deque contains the following lines of code:
\begin{lstlisting}
std::atomic_thread_fence(std::memory_order_release);
_bottom[p].data.store(b + 1, std::memory_order_relaxed);
\end{lstlisting}
If memory\_order\_relaxed is replaced here by memory\_order\_release, the sanitizer stops detecting the data
races. This might indicate a false positive related to the usage of std::atomic\_thread\_fence. It is worth
noting that Taskflow unit tests and examples pass with the thread sanitizer even though std::atomic\_thread\_fence
is used.

An example of a work-stealing deque implementation that does not use std::atomic\_thread\_fence can be found
in Google Filament \cite{Filament}. When using the implementation from Google Filament, the thread sanitizer does
not detect data races in the suggested solution.

Concurrent push and pop operations are not allowed in most implementations of the work-stealing deque. To ensure
that there are no concurrent push and pop operations, mappings from thread ID to queue indices are typically used.
When a task is pushed to or popped from a queue, the correct queue is usually found using the current thread ID.
Unlike this typical approach, the solution suggested in this paper uses a thread-local variable to find the
correct task queue. It makes the solution not header-only, but this will not matter once modules from the C++20
standard are used. Unfortunately, at the time of writing this paper, there seems to be not enough compiler
support to present the suggested solution in the form of a cross-platform module.

\subsection{Task graphs}
To run task graphs, simple wrappers over an std::function\textless void()\textgreater are used. Each wrapper stores references to
successor tasks and the number of uncompleted predecessor tasks. When the thread pool executes a task, it first
executes the wrapped function. Then, for each successor task, it decrements the number of uncompleted
predecessor tasks. One of the successor tasks, for which the number of uncompleted predecessor tasks becomes equal
to zero, is then executed on the same worker thread. Other successor tasks, for which the number of uncompleted
predecessor tasks becomes equal to zero, are submitted to the same thread pool instance for execution.

\section{Benchmarks}
In this section, we provide some information about benchmark results for the suggested solution.

Due to its minimalism and simplicity, the suggested solution seems to demonstrate good CPU performance. A few
simple benchmarks have been used to compare the CPU performance of the suggested solution with Taskflow. For example,
a simple recursive function to calculate Fibonacci numbers without memoization, taken from Taskflow examples, can be
used to evaluate performance when running a large number of tasks. The charts below demonstrate the results
at the time of writing this paper:

\begin{center}
\includegraphics[scale=0.6]{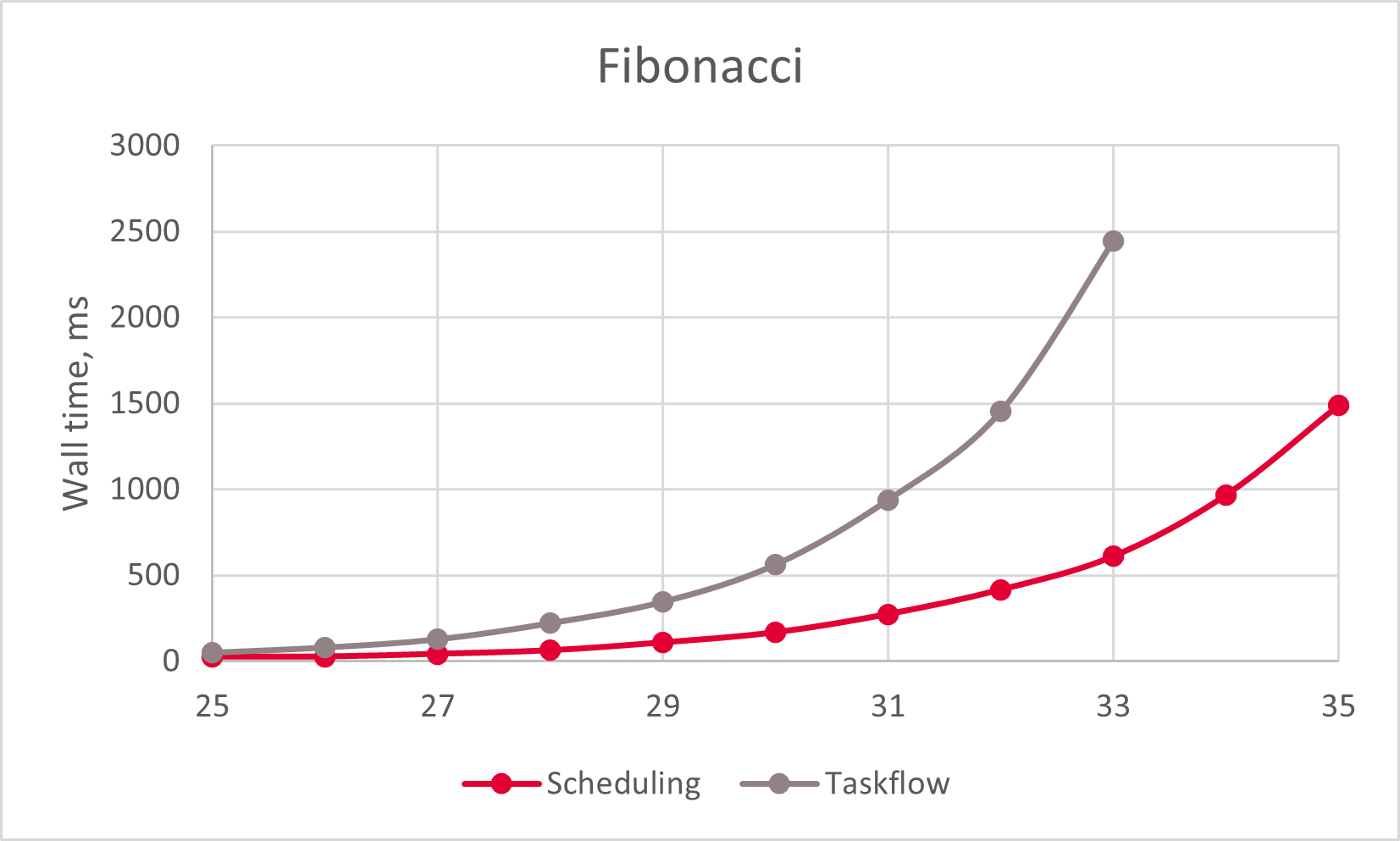}

Fig. 1. Wall time
\end{center}

\begin{center}
\includegraphics[scale=0.6]{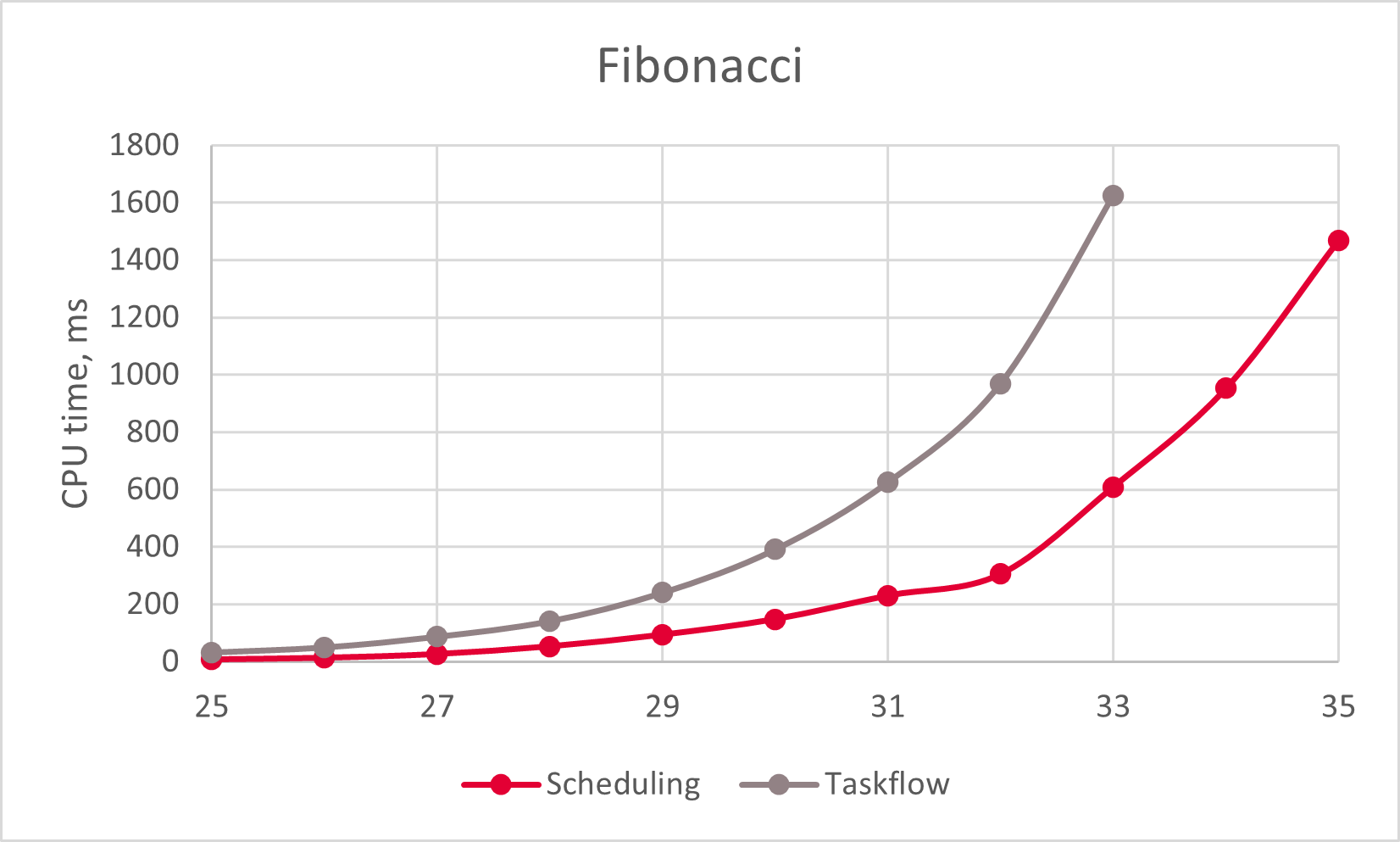}

Fig. 2. CPU time
\end{center}
These charts indicate that in simple use cases, the CPU performance of the suggested solution can be quite comparable to
Taskflow. For more benchmark results, see https://github.com/dpuyda/scheduling.

\section{Usage}
In this section, we give an idea of how to use the solution suggested in this paper. More details about how to use the
suggested solution can be found at https://github.com/dpuyda/scheduling.

\subsection{Async tasks}
Here, we briefly describe how to run async tasks using the suggested solution.

To run an async task, first, create a ThreadPool instance. For example:
\begin{lstlisting}
#include "scheduling/scheduling.hpp"
...
scheduling::ThreadPool thread_pool;
\end{lstlisting}
In the constructor, the ThreadPool class creates several worker threads that will be running in the background
until the instance is destroyed. As an optional argument, the constructor of the ThreadPool class accepts the
number of threads to create. By default, the number of threads is equal to std::thread::hardware \_concurrency().

When the ThreadPool instance is created, submit a task. For example:
\begin{lstlisting}
thread_pool.Submit([] {
  std::this_thread::sleep_for(std::chrono::seconds(1));
  std::cout << "Completed\n";
});
\end{lstlisting}
A task is a function that does not accept arguments and returns void. Use lambda captures to pass input and
output arguments to the task if needed. Eventually, the task will be executed on one of the worker threads owned
by the ThreadPool instance.

\subsection{Task graphs}
Here, we briefly describe how to run task graphs using the suggested solution.

A task graph is a collection of tasks and dependencies between them. Dependencies between tasks define the
order in which the tasks should be executed. Consider a simple illustrative example of a task graph. Assume that
we want to calculate the arithmetic expression $(a + b) * (c + d)$, and that each operation in this expression
(including getting the values of $a$, $b$, $c$, and $d$) takes time. To optimize execution time, we can start by
getting the values of $a$, $b$, $c$, and $d$ in parallel. Then, once we know the values of $a$ and $b$, we can
start calculating the value of $a + b$, and once we know the values of $c$ and $d$, we can start calculating the
value of $c + d$. The values of $a + b$ and $c + d$ can be calculated in parallel. Once $a + b$ and $c + d$ are
calculated, we can start calculating $(a + b) * (c + d)$.

The code snippets below illustrate how to execute the above task graph using the suggested solution. To define a
task graph, create an iterable collection of Task instances. For example:
\begin{lstlisting}
#include "scheduling/scheduling.hpp"
...
std::vector<scheduling::Task> tasks;
\end{lstlisting}
Add elements to tasks. For example, let us add tasks to calculate the value of $(a + b) * (c + d)$
asynchronously. First, add tasks to get the values of $a$, $b$, $c$ and $d$:
\begin{lstlisting}
int a, b, c, d;

auto& get_a = tasks.emplace_back([&] {
  std::this_thread::sleep_for(std::chrono::seconds(1));
  a = 1;
});

auto& get_b = tasks.emplace_back([&] {
  std::this_thread::sleep_for(std::chrono::seconds(1));
  b = 2;
});

auto& get_c = tasks.emplace_back([&] {
  std::this_thread::sleep_for(std::chrono::seconds(1));
  c = 3;
});

auto& get_d = tasks.emplace_back([&] {
  std::this_thread::sleep_for(std::chrono::seconds(1));
  d = 4;
});
\end{lstlisting}

Next, add tasks to calculate $a + b$ and $c + d$:
\begin{lstlisting}
int sum_ab, sum_cd;

auto& get_sum_ab = tasks.emplace_back([&] {
  std::this_thread::sleep_for(std::chrono::seconds(1));
  sum_ab = a + b;
});

auto& get_sum_cd = tasks.emplace_back([&] {
  std::this_thread::sleep_for(std::chrono::seconds(1));
  sum_cd = c + d;
});
\end{lstlisting}

Finally, add the task to calculate the product $(a + b) * (c + d)$:
\begin{lstlisting}
int product;

auto& get_product = tasks.emplace_back([&] {
  std::this_thread::sleep_for(std::chrono::seconds(1));
  product = sum_ab * sum_cd;
});
\end{lstlisting}

When all tasks are added, define task dependencies. The task get\_sum\_ab should be executed after
get\_a and get\_b. Similarly, the task get\_sum\_cd should be executed after get\_c and get\_d.
Finally, the task get\_product should be executed after get\_sum\_ab and get\_sum\_cd:
\begin{lstlisting}
get_sum_ab.Succeed(&get_a, &get_b);
get_sum_cd.Succeed(&get_c, &get_d);
get_product.Succeed(&get_sum_ab, &get_sum_cd);
\end{lstlisting}

When dependencies between tasks are defined, create a ThreadPool instance and submit the task
graph for execution:
\begin{lstlisting}
scheduling::ThreadPool thread_pool;
thread_pool.Submit(tasks);
\end{lstlisting}

See https://github.com/dpuyda/scheduling for more details about how to run async tasks and task
graphs using the suggested solution.

\section{License}
The suggested solution is licensed under the MIT License. It includes code from Google Filament, which is licensed under
the Apache License 2.0, and code from Taskflow, which is licensed under the MIT License.

\section*{Acknowledgements}
The author would like to thank Jason Turner and Tsung-Wei Huang for useful discussions. The author
also would like to thank Chunel Feng for bringing up an interesting example that helped to improve
the implementation of the suggested solution.

\end{document}